\documentclass[11pt,preprintnumbers,aps,amssymb,nofootinbib,amsmath,superscriptaddress]
{revtex4}
\usepackage{epsfig,epsf}
\usepackage{bm} 
\newcommand{\beq}{\begin{equation}}
\newcommand{\beql}[1]{\begin{equation}\label{#1}}
\newcommand{\eeq}{\end{equation}}
\newcommand{\eq}[1]{(\ref{#1})}
\newcommand{\fig}[1]{Fig.~\ref{#1}}
\renewcommand{\sec}[1]{Sec.~\ref{#1}}
\newcounter{topiccounter}
\renewcommand{\b}[1]{{\bm #1}} 
\newcommand{\unit}[1]{\hat {{\bm #1}}} 

\newcommand{\e}{\varepsilon}

\begin{document}

\title{
Electromagnetic radiation by quark-gluon plasma in magnetic field}

\author{Kirill Tuchin}

\affiliation{
Department of Physics and Astronomy, Iowa State University, Ames, IA 50011}

\date{\today}


\begin{abstract}

The electromagnetic radiation by quark-gluon plasma in strong magnetic field is calculated. The contributing processes  are  synchrotron radiation and one--photon annihilation. It is shown that in relativistic heavy--ion collisions at RHIC and LHC synchrotron radiation dominates over the annihilation. Moreover, it  constitutes a significant part of all  photons produced  by the plasma at low transverse momenta; its magnitude depends on the plasma temperature and the magnetic field strength.  Electromagnetic radiation in magnetic field is probably the missing piece that resolves a discrepancy between the theoretical models and the experimental data. It is argued that electromagnetic radiation  increases with the magnetic  field strength and plasma temperature.

\end{abstract}

\maketitle

\section{Introduction}\label{sec:intr}

It has been known for a long time that strong magnetic fields can be generated in heavy-ion collisions. Recent calculations \cite{Kharzeev:2007jp,Skokov:2009qp,Voronyuk:2011jd,Deng:2012pc,Bzdak:2011yy}  confirm this wisdom and predict, using essentially classical 
electrodynamics, that the magnetic field shortly after the collision reaches $10^{18}-10^{19}$~G, which by far exceeds the critical value $B_c=4.41\cdot 10^{13}$~G for electrons, known also as the Schwinger field. 
However, it was only recently appreciated that such fields may have great phenomenological significance. There are two observations leading to this conclusion: (i) phenomenological models  suggest that the Quark Gluon Plasma (QGP) is formed after a very short time after the heavy--ion collision -- on the order of a few tenth of a fm/c; (ii) 
the relaxation time of  magnetic field in the presence of QGP is proportional to the plasma electric conductivity due to the induction of electric Foucault currents by the time-dependent magnetic field \cite{Tuchin:2010vs}. The relaxation time of magnetic field is estimated to be about 1--2 few fm/c \cite{Tuchin:2010vs}. Therefore, magnetic field has a profound influence on all aspects of the physics of relativistic heavy--ion collisons. In particular, it was argued in \cite{Tuchin:2010vs} that magnetic field induces energy loss by fast quarks and charged leptons via the synchrotron radiation and  polarization of the fermion spectra. It contributes to enhancement of dilepton production  at low invariant masses \cite{Tuchin:2010gx} and enhances the azimuthal anisotropy of the quark-gluon plasma (QGP) \cite{Mohapatra:2011ku,Tuchin:2011jw}. 
It causes  dissociation of the bound states, particularly charmonia,  via ionization  \cite{Marasinghe:2011bt,Tuchin:2011cg}. 
Additionally,  magnetic field drives the Chiral Magnetic Effect (CME) \cite{Kharzeev:2004ey,Kharzeev:2007jp,Kharzeev:2007tn,Fukushima:2008xe,Kharzeev:2009fn}, which is the generation of an electric field parallel to the magnetic one via the axial anomaly in the hot nuclear matter. Additionally, the effect of the magnetic field on the QCD phase diagram was studied using model calculations \cite{Mizher:2010zb,Fraga:2008um,Gatto:2010pt,Gatto:2010qs,Osipov:2007je,Kashiwa:2011js,Johnson:2008vna,Kanemura:1997vi,Alexandre:2000yf,Agasian:2008tb,Preis:2010cq,Gusynin:1995nb,Miransky:2002rp,Boomsma:2009yk,Shushpanov:1997sf,Cohen:2007bt,Agasian:2001hv,D'Elia:2010nq} and lattice simulations \cite{Cea:2005td,Cea:2007yv,Cea:2002wx,D'Elia:2010nq,Bali:2011qj}.  It is also argued in Appendix,  that at RHIC, at early times after heavy ion collision, about 3\% of energy density of plasma resides in the magnetic field, while at LHC, this fraction reaches as much as 40\%.

This paper addresses the problem of photon radiation by quarks and antiquarks of QGP moving in external magnetic field. This radiation originates from two sources: (i) synchrotron radiation and (ii) quark and antiquark annihilation. QGP is transparent to the emitted electromagnetic radiation because its absorption coefficient  is suppressed by $\alpha^2$. Thus,  QGP is shinning in magnetic field. The main goal of this paper is to calculate the spectrum and angular distribution of this radiation. In strong magnetic field it is essential to account for quantization of fermion spectra. Indeed, spacing between the Landau levels is of the order $\sqrt{e B}$, while their thermal width is of the order $T$.  Spectrum quantization is negligible only if $eB\ll T^2$ which is barely the case at RHIC and certainly not the case at LHC (at least during the first few fm's of the evolution). Fermion spectrum quantization is important not only for hard and electromagnetic probes but also for the bulk properties of QGP.

The presentation is structured as follows.  In \sec{sec:synch} the spectrum and angular distribution of synchrotron radiation by QGP is calculated in the ideal gas approximation and compared to the experimental data. The results exhibited  in \fig{tot-synch},\ref{synch-err} indicate that photon radiation in magnetic field gives a significant contribution to the total photon yield. This contribution seems to be large enough to account for the discrepancy between the model calculations assuming no magnetic field and the experimental data at RHIC \cite{Adare:2008ab}.\footnote{Another possible explanation has been recently suggested in \cite{Chiu:2012ij}.} Moreover, the electromagnetic radiation rapidly increases with $B$ and $T$ suggesting that it plays even more important role at LHC. In \sec{sec:annih} the photon spectrum emitted in pair annihilation is calculated and it is shown that it is small as compared to synchrotron contribution. A possible way to ascertain existence of  synchrotron radiation is discussed in \sec{sec:concl}.

\section{Synchrotron radiation}\label{sec:synch}

Motion of charged fermions in external magnetic field, which we will approximately treat as spatially homogeneous,  is quasi-classical in the field  direction and quantized in the \emph{reaction plane}, which is  perpendicular to the magnetic field and span by the impact parameter and the heavy ion collision axis.   In high energy physics  one usually distinguishes the \emph{transverse plane}, which is  perpendicular to the collision axis and span by the magnetic field and the impact parameter. The notation is adopted in which three-vectors are discriminated by the bold face and their  component along the field direction by the plain face. Momentum projections onto the transverse plane are denoted by subscript $\bot$.

In the configuration space, charged fermions move along spiral trajectories with the symmetry axis aligned with the field direction. Synchrotron radiation is a process of photon $\gamma$ radiation by a fermion $f$ with electric charge $e_f=z_f e$  in external magnetic field $B$:
\beql{srad}
f(e_f,j,p)\to f(e_f,k,q)+\gamma(\b k)\,,
\eeq
where $\b k$ is the photon momentum, $p,q$ are the momentum components along the magnetic field direction and 
indicies $j,k=0,1,2,\ldots$  label the discrete Landau levels in the reaction plane.  
The Landau levels  are given  by
\begin{align}\label{mass-shell}
\e_j= \sqrt{m^2+p^2+2j e_f  B}\,,\quad \e_k= \sqrt{m^2+q^2+2ke_f  B}\,,
\end{align}
In the constant magnetic field only momentum component along the field direction is conserved. Thus, the  conservation laws  for synchrotron radiation read
\begin{align}\label{conserv}
\e_j= \omega +\e_k\,,\quad p=q+\omega\cos\theta\,,
\end{align}
where $\omega$ is the photon energy and $\theta$ is the photon emission angle with respect to the magnetic field. Intensity of the synchrotron radiation was derived in \cite{Sokolov:1968a}. In \cite{Herold:1982a,Harding:1987a,Latal:1986a,Baring:1988a}  it was thoroughly investigated as a possible mechanism for $\gamma$-ray bursts. In particular, synchrotron radiation in electromagnetic plasmas was calculated. 
Spectral intensity of angular distribution of synchrotron radiation by a fermion in the $j$'th Landau state  is given by 
\begin{align}\label{intj}
\frac{dI^j}{d\omega d\Omega}=\sum_f\frac{z_f^2 \alpha}{\pi}\omega^2\sum_{k=0}^j\Gamma_{jk}\left\{
|\mathcal{M}_\bot|^2+|\mathcal{M}_\parallel|^2\right\}\,\delta(\omega-\e_j+\e_k)
\end{align}
where $\Gamma_{jk} = (1+\delta_{j0})(1+\delta_{k0})$ accounts for the double degeneration of all Landau levels except the ground one.  The squares of matrix elements $\mathcal{M}$, which  appear in \eq{intj}, corresponding to photon polarization perpendicular and parallel to the magnetic field are given by, respectively,
\begin{align}
4\e_j\e_k|\mathcal{M}_\bot|^2= &(\e_j\e_k-pq-m^2)[I_{j,k-1}^2+I_{j-1,k}^2]+2\sqrt{2j e_f   B}\sqrt{2k e_f   B}[I_{j,k-1}I_{j-1,k}]\label{Mmat1}\,.\\
 4\e_j\e_k|\mathcal{M}_\parallel|^2=& \cos^2\theta\big\{ ( \e_j\e_k-pq-m^2)[I_{j,k-1}^2+I_{j-1,k}^2]-2\sqrt{2je_f   B}\sqrt{2k e_f   B}[I_{j,k-1}I_{j-1,k}]\big\}\nonumber\\
 &-2\cos\theta\sin\theta\big\{ p\sqrt{2 k e_f   B}[I_{j-1,k}I_{j-1,k-1}+I_{j,k-1}I_{j,k}]\nonumber\\
  &  +
 q\sqrt{2je_f   B}[I_{j,k}I_{j-1,k}+I_{j-1,k-1}I_{j,k-1}]\big\}\nonumber\\
 &+ \sin^2\theta\big\{ (\e_j\e_k+pq-m^2)[I_{j-1,k-1}^2+I_{j,k}^2]+2\sqrt{2je_f   B}\sqrt{2ke_f   B}(I_{j-1,k-1}I_{j,k})\big\}\,, \label{Mmat2}
\end{align}
where for $j\ge k$, 
\beql{ijk} 
I_{j,k}\equiv I_{j,k}(x)= (-1)^{j-k}\sqrt{\frac{k!}{j!}}e^{-\frac{x}{2}}x^{\frac{j-k}{2}}L_k^{j-k}(x).
\eeq
and $I_{j,k}(x)= I_{k,j}(x)$ when $k>j$. ($I_{j,-1}$ are identically zero).
The functions $L_k^{j-k}(x)$ are the generalized Laguerre polynomials. Their argument is 
\beql{xa}
x=\frac{\omega^2}{2e_f   B}\sin^2\theta\,.
\eeq

Angular distribution of radiation is obtained by  integrating over the photon energies and remembering that $\e_k$ also depends on $\omega$ by virtue of \eq{mass-shell} and \eq{conserv}:
\begin{align}\label{dIdO}
\frac{dI^j}{d\Omega}=\sum_f\frac{z_f  ^2 \alpha}{\pi}\sum_{k=0}^j \frac{\omega^*(\e_j-\omega^*)}{\e_j-p\cos\theta -\omega^*\sin^2\theta}\Gamma_{jk} \left\{
|\mathcal{M}_\bot|^2+|\mathcal{M}_\parallel|^2\right\}\,,
\end{align}
where photon energy $\omega$ is fixed to be 
\beql{omega*}
\omega^* = \frac{1}{\sin^{2}\theta}\left\{(\e_j-p\cos\theta)-\big[(\e_j-p\cos\theta)^2-2e_f   B(j-k)\sin^2\theta\big]^{1/2}\right\}\,.
\eeq

In the context  of heavy-ion collisions the relevant  observable is the differential photon spectrum. For ideal plasma in equilibrium each quark flavor gives the following contribution to the photon spectrum:
\begin{align}\label{def-sp}
\frac{dN^\text{synch}}{dt d\Omega d\omega} =\sum_f \int_{-\infty}^\infty dp \frac{e_f  B (2N_c) V}{2\pi^2} \sum_{j=0}^\infty\sum_{k=0}^j \frac{dI^j}{\omega d\omega d\Omega}(2-\delta_{j,0}) f(\e_j)[1-f(\e_k)]\,,
\end{align}
where $2N_c$ accounts for quarks and antiquarks each of  $N_c$ possible colors, and $(2-\delta_{j,0})$ sums over the initial quark spin. Index $f$ indicates different quark flavors. $V$ stands for the plasma volume. 
The statistical factor $f(\e)$ is
\beql{stat}
f(\e)= \frac{1}{e^{\e/T}+1}\,.
\eeq
The $\delta$-function appearing in \eq{intj} can be re-written 
using \eq{mass-shell} and \eq{conserv}  as 
\beql{re1}
\delta(\omega-\e_j+\e_k) = \sum_{\pm}\frac{\delta(p-p^*_\pm)}{\big| \frac{p}{\e_j}-\frac{q}{\e_k}\big|}\,,
\eeq
where 
\begin{align}\label{p*}
p^*_\pm= &\bigg\{\cos\theta (m_j^2-m_k^2+\omega^2\sin^2\theta)\nonumber\\
&\pm \sqrt{[(m_j+m_k)^2-\omega^2\sin^2\theta][(m_j-m_k)^2-\omega^2\sin^2\theta]}\,\bigg\}/(2\omega\sin^2\theta)\,.
\end{align}
The following convenient notation was introduced:
\beql{mm}
m_j^2= m^2+2je_f  B\,,\quad m_k^2= m^2+2ke_f  B\,.
\eeq
The physical meaning of \eq{p*} is that synchrotron radiation of a  photon with energy $\omega$ at angle $\theta$ by a fermion undergoing transition from $j$'th to $k$'th Landau level is possible only if the initial quark momentum along the field direction equals $p^*_\pm$.  

Another consequence of the conservation laws  \eq{conserv} is that 
for a given $j$ and $k$ the photon energy cannot exceed a certain maximal value that will be denoted by $\omega_{s,jk}$.  Indeed, inspection of \eq{p*} reveals that this equation has a real solution only in two cases 
\beql{cases}
\text{(i)}\,\, m_j-m_k\ge\omega \sin\theta\,,\quad \text{or}\quad \text{(ii)}\,\, m_j+m_k\le \omega \sin\theta\,.
\eeq
The first case is relevant for the synchrotron radiation while the second one for the one-photon pair annihilation as discussed in the next section. Accordingly, allowed photon energies in the $j\to k$ transition satisfy  
\beql{omegac}
\omega\le \omega_{s,jk} \equiv \frac{m_j-m_k}{\sin\theta}= \frac{\sqrt{m^2+2je_f   B}-\sqrt{m^2+2ke_f  B}}{\sin\theta}\,.
\eeq 
No synchrotron radiation is possible for  $\omega>\omega_{s,jk}$. In particular, when $j=k$, $\omega_{s,jk}=0$, i.e.\ no photon is emitted, which is also evident in   \eq{omega*}.  The reason is clearly seen in the frame where $p=0$: since $\e_j\ge \e_k$, constraints \eq{mass-shell} and \eq{conserv} hold only if $\omega=0$. 

Substitution of \eq{intj} into \eq{def-sp} yields the spectral distribution of the synchrotron radiation rate per unit volume
\begin{align}\label{spec1}
\frac{dN^\text{synch}}{Vdt d\Omega d\omega}= \sum_f\frac{2N_cz_f^2\alpha}{\pi^3}e_f  B 
\sum_{j=0}^\infty \sum_{k=0}^j \omega (1+\delta_{k0}) \,\vartheta(\omega_{s,ij}-\omega)
\int dp \sum_\pm\frac{\delta(p-p^*_\pm)}
{\big| \frac{p}{\e_j}-\frac{q}{\e_k}\big|}
&\nonumber\\
\times\left\{
|\mathcal{M}_\bot|^2+|\mathcal{M}_\parallel|^2\right\} f(\e_j)[1-f(\e_k)] \,,&
\end{align} 
where $\vartheta$ is the step-function. 

The natural variables to study the synchrotron radiation are the photon energy $\omega$ and its emission angle $\theta$ with respect to the magnetic field. However, in high energy physics particle spectra are traditionally presented in terms of  rapidity $y$ (which for photons is equivalent to pseudo-rapidity)  and transverse momentum $k_\bot$. $k_\bot$ is a projection of three-momentum $\b k$ onto the transverse plane. These variables are not convenient to study electromagnetic processes in external magnetic field. In particular, they conceal the azimuthal symmetry with respect to the magnetic field direction. To change variables, let $z$ be the collision axis and $\unit y$ be the direction of the magnetic field. In spherical coordinates photon momentum is given by $\b k =\omega( \sin\alpha\cos\phi\unit x+\sin\alpha\sin\phi\unit y+\cos\alpha\unit z)$, where $\alpha$ and $\phi$ are the polar and azimuthal angles with respect to $z$-axis. The plane $xz$ is  the reaction plane.  By definition, $\unit k\cdot \unit y = \cos\theta$ implying that $\cos\theta= \sin\alpha\sin\phi$. Thus, 
\beql{coor1}
k_\bot=\sqrt{ k_x^2+k_y^2} = \frac{\omega\cos\theta}{\sin\phi}\,,\quad y= -\ln\tan\frac{\alpha}{2}\,.
\eeq  
The second of these equations is the definition of (pseudo)-rapidity. Inverting \eq{coor1} yields
\beql{coor2}
\omega= k_\bot \cosh y\,,\quad \cos\theta= \frac{\sin\phi}{\cosh y}\,.
\eeq
Because $dy=dk_z/\omega$ the photon multiplicity in a unit volume per unit time reads
\beql{mult2}
\frac{dN^\text{synch}}{dVdt\,d^2k_\bot dy}=\omega\frac{dN^\text{synch}}{dVdt\,d^3k}= \frac{dN^\text{synch}}{dVdt\,\omega d\omega d\Omega}
\eeq

\begin{figure}[ht]
\begin{tabular}{cc}
      \includegraphics[height=5cm]{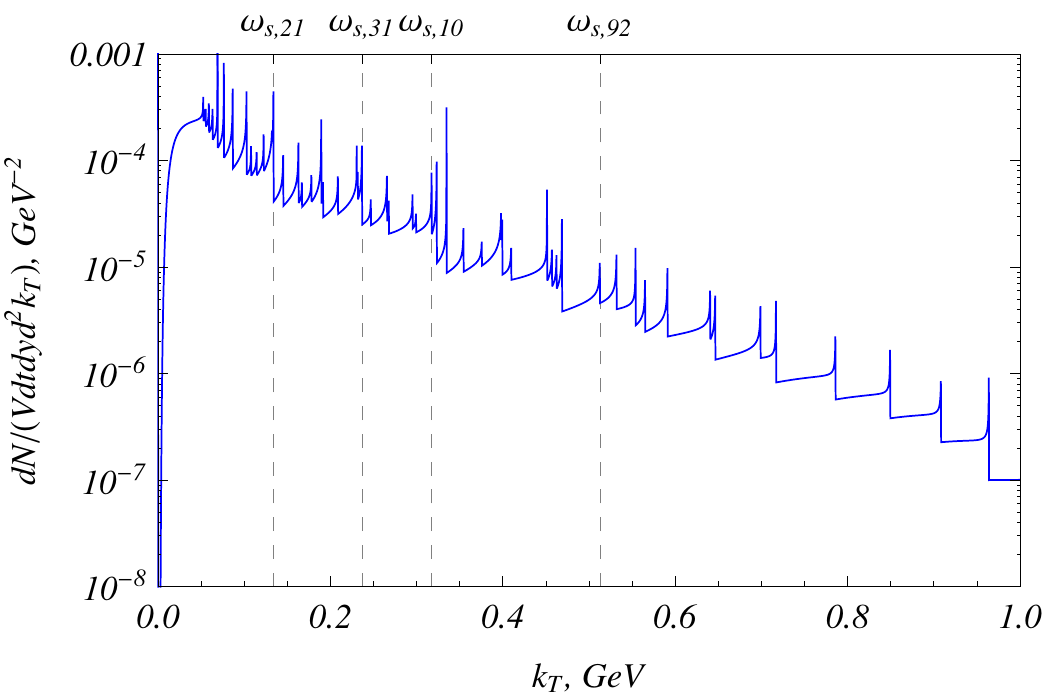} &
      \includegraphics[height=5cm]{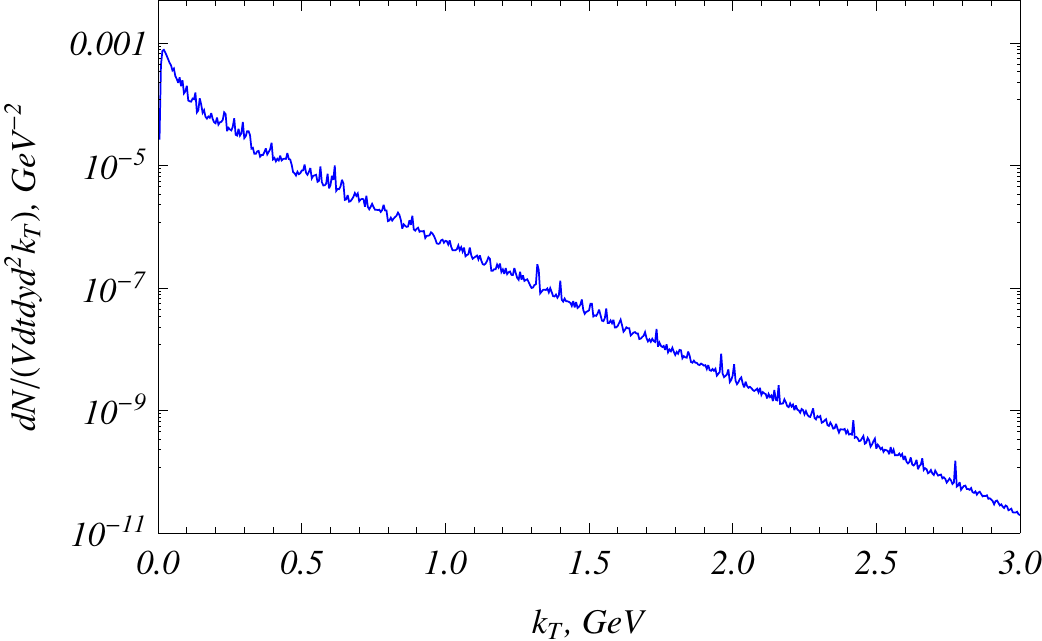}\\
      $(a)$ & $(b)$ 
      \end{tabular}
  \caption{Spectrum of synchrotron radiation by $u$ quarks at $eB=m_\pi^2$, $y=0$, $\phi=\pi/3$: (a) contribution of 10 lowest Landau levels $j\le 10$; several cutoff frequencies are indicated; (b) summed over all Landau levels. $m_u=3$~MeV, $T=200$~MeV.}
\label{synch}
\end{figure}
\begin{figure}[ht]
      \includegraphics[height=5cm]{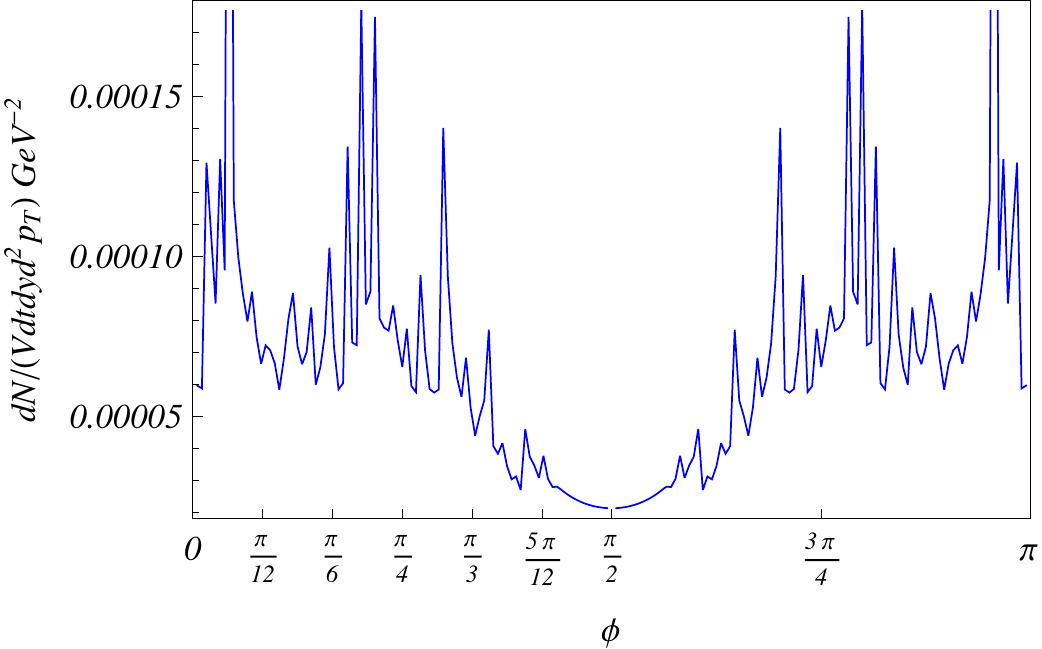} 
  \caption{Azimuthal distribution of synchrotron radiation by $u$-quarks at  $k_\bot=0.2$~GeV, $eB=m_\pi^2$, $y=0$. $m_u=3$~MeV.}
\label{synch2}
\end{figure}

\fig{synch} displays the spectrum of synchrotron radiation by $u$ quarks as a function of $k_\bot$ at fixed $\phi$. At midrapidity $y=0$ \eq{coor2} implies that $k_\bot=\omega$.  Contribution of $d$ and $s$ quarks is qualitatively similar. At $eB\gg m^2$, quark masses do not  affect the spectrum much.  The main difference stems from the difference in electric charge. In panel (a) only the contributions of the first ten Landau levels are displayed. The cutoff frequencies $\omega_{s,jk}$ can be clearly seen and some of them are indicated on the plot for convenience. The azimuthal distribution is shown in \fig{synch2}. Note, that  at midrapidity $\phi= \pi/2-\theta$. Therefore, the figure indicates that photon production in the direction of magnetic field (at $\phi= \pi/2$) is suppressed. More photons are produced in the direction of the reaction plane $\phi=0$. This results in the ellipticity of the photon spectrum that translates into the positive ``elliptic flow" coefficient $v_2$. It should be noted, that the classical synchrotron radiation has a similar angular  distribution. 

\begin{figure}[ht]
\begin{tabular}{cc}
      \includegraphics[height=5cm]{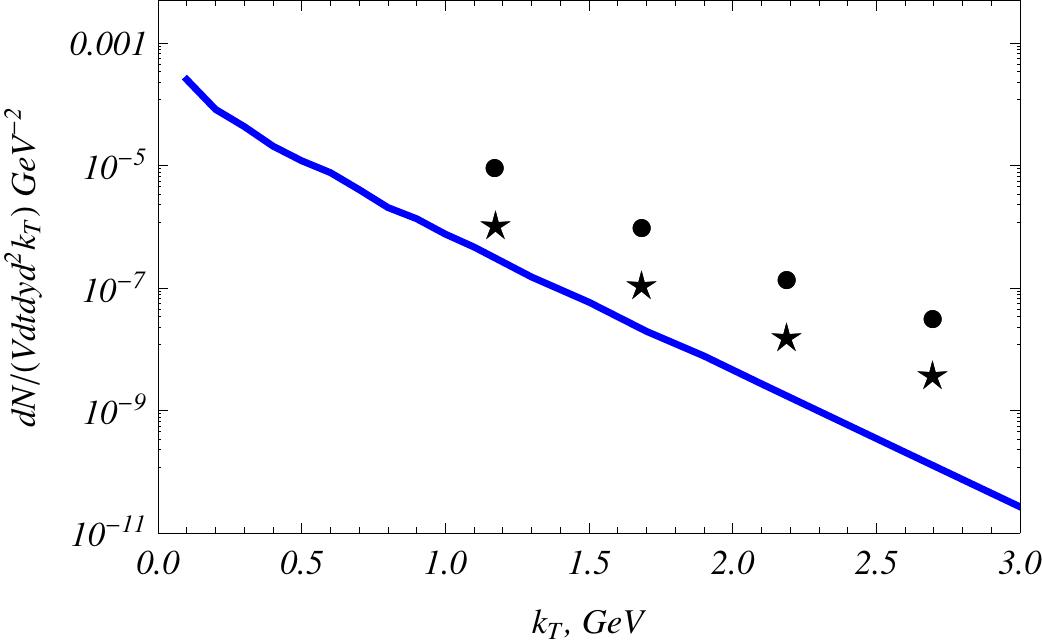} &
      \includegraphics[height=5.cm]{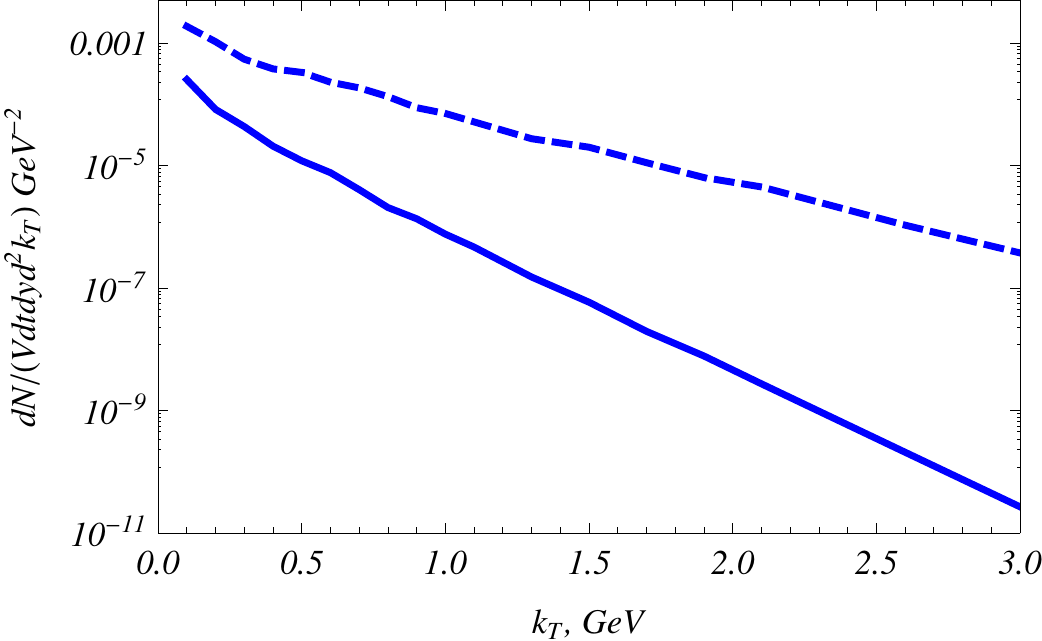}\\
      $(a)$ & $(b)$ 
      \end{tabular}
  \caption{ Azimuthal average of the synchrotron radiation spectrum of $u$,$d$,$s$ quarks and their corresponding antiquarks.  (a) $eB=m_\pi^2$, $y=0$ compared to the experimental data from \cite{Adare:2008ab} divided by $Vt=25\pi$~fm$^4$ (dots) and 
 $Vt=9\times 25\pi$~fm$^4$ (stars), (b)  $eB=m_\pi^2$, $T=200$~MeV, $y=0$ (solid line) compared to $eB=15m_\pi^2$, $T=400$~MeV, $y=0$ (dashed line). $m_u=3$~MeV, $m_d=5$~MeV, $m_s=92$~MeV.}
\label{tot-synch}
\end{figure}

In order to compare the photon spectrum produced by synchrotron radiation to the photon spectrum measured in heavy-ion collisions, the $u$, $d$ and $s$ quarks contributions were summed up. Furthermore, the experimental data from \cite{Adare:2008ab} was divided by $Vt$, where $t$ is the magnetic field relaxation time.  The volume of the plasma can be estimated as $V=\pi R^2t$ with $R\approx 5$~fm being the nuclear radius. Therefore, 
\beql{mult4}
\frac{dN^\gamma_\text{exp}}{dVdt\, d^2k_\bot dy}=\frac{dN^\gamma_\text{exp}}{d^2k_\bot dy}\,\frac{1}{\pi R^2t^2}=\frac{dN^\gamma_\text{exp}}{d^2k_\bot dy}\,\left(\frac{\text{GeV}}{14.9}\right)^4\,\left( \frac{1\,\text{fm}}{t}\right)^2\,.
\eeq
 The results are plotted in \fig{tot-synch}. In panel (a) it is seen that synchrotron radiation gives a significant contribution to the photon production in heavy-ion collisions at RHIC energy. This contribution is larger at small transverse momenta. This may explain enhancement of  photon production observed in \cite{Adare:2008ab}. Panel (b) indicates the increase of the photon spectrum produced by the synchrotron radiation mechanism at the LHC energy. This increase is due to enhancement of the magnetic field strength, but mostly because of  increase of plasma temperature.  This qualitative features can be better understood by considering the limiting cases of low and high photon energies.

\subsection{Low photon energy}

The low energy  part of the photon spectrum satisfies the condition  $\omega\ll \sqrt{e_fB}$. The corresponding initial quark momentum component along the field $p$ and energy $\e_j$ follow from \eq{p*} and \eq{mass-shell} and are given by
\beql{papprox}
 p_\pm^*\approx \frac{(j-k)e_fB(\cos\theta\pm 1)}{\omega\sin^2\theta}+\mathcal{O}(\omega)\,,\qquad \e_j\approx |p_\pm^*|+\mathcal{O}(\omega)\,.
\eeq
Evidently, $\e_j\gg eB$. In practice, the magnetic field strength satisfies $\sqrt{eB}\gtrsim T$, so that $\e_j\gg T$. Therefore, synchrotron radiation is dominated by fermion transitions from low Landau levels due to the statistical factors appearing in \eq{def-sp}. 

For a qualitative discussion it is sufficient to consider the $1\to 0$ transition. In this case the matrix elements \eq{Mmat1} and \eq{Mmat2} read
\beql{m10}
|\mathcal{M}^{1,0}|^2=\frac{1}{2\e_1\e_0}\left\{ I_{1,0}^2(\e_1\e_0 -pq\cos^2\theta-m^2)+\cos\theta\sin\theta q\sqrt{2e_f   B}I_{1,0}I_{0,0}\right\}\,.
\eeq
Assuming that the field strength is supercritical, i.e.\  $e_f   B\gg m^2$, but keeping all powers of  $\omega$ (for future reference) \eq{p*} reduces to 
\beql{p*sc}
p_\pm^*\approx \frac{1}{2\omega \sin^2\theta}\left\{ 2e_f  B(\cos\theta\pm 1)+\omega^2 \sin^2\theta (\cos\theta\mp 1)\right\}\,.
\eeq
Furthermore, using the conservation laws \eq{conserv} we obtain in this approximation
\begin{align}
\e_{1\pm}= &\frac{1}{2\omega \sin^2\theta}\left| 2e_f   B(\cos\theta\pm 1)-\omega^2 \sin^2\theta (\cos\theta \mp 1)\right|\label{en-mom1}\,,\\
q_\pm=& \frac{1}{2\omega \sin^2\theta}(2e_f  B-\omega^2\sin^2\theta)(\cos\theta\pm 1)\,,
\label{en-mom2}\\
\e_{0\pm}=&|q|\,.\label{en-mom3}
\end{align}
The values of the non-vanishing matrix elements $I_{j,k}$ defined by \eq{ijk} are
\begin{align}\label{is}
I_{1,0}(x)&= -x^{1/2}e^{-x/2}\,,\qquad I_{0,0}(x)= e^{-x/2}\,.
\end{align}
For $j=1$, $k=0$ we write using \eq{omegac} $\omega_{s,10}= \sqrt{2e_f  B}/\sin\theta$. Then \eq{xa} implies $x=\omega^2/\omega_{s,10}^2$. 
Substituting \eq{p*sc}--\eq{is} into \eq{m10} gives
\begin{align}
|\mathcal{M}^{1,0}_\pm|^2= \frac{1}{2}xe^{-x}\left[ 1-
\frac{\cos\theta(1+x)\pm (1-x)}{\cos\theta(1-x)\pm (1+x)}\cos^2\theta
-\frac{2(1-x)\cos\theta\sin^2\theta}{\cos\theta(1-x)\pm (1+x)} 
\right]\,.
\end{align}
According to \eq{spec1} the contribution of the $1\to 0$ transition to the synchrotron radiation reads 
\begin{align}\label{ap1}
\frac{dN^{\text{synch},10}}{Vdt d\Omega d\omega }= \sum_f\frac{2N_cz_f^2\alpha}{\pi}\omega \Gamma\frac{e_f  B}{2\pi^2}
\sum_\pm  f(\e_1)[1-f(\e_0)]|\mathcal{M}^{1,0}_\pm|^2\,&\nonumber\\
\times\frac{(1-x)\cos\theta\pm (1+x)}{-2x(\cos\theta\mp 1)}\,\vartheta(\omega_{s,10}-\omega)\,.&
\end{align}
Consider radiation spectrum at $\theta=\pi/2$, i.e.\ perpendicular to the magnetic field. The spectrum increases with $x$ and reaches maximum at $x=1$. Since  $x=\omega^2/(2e_f   B)$, spectrum decreases with increase of $B$ at fixed $\omega$. This feature holds at low $x$ part of the spectrum for other emission angles and even for transitions form higher excited states. However, at high energies, it is no longer possible to approximate the spectrum by the contribution of a few low Landau levels. In that case the typical values of quantum numbers are $j,k\gg 1$. For example, to achieve the numerical accuracy of 5\%, sum over $j$ must run up to a certain $j_\text{max}$. Some values of $j_\text{max}$ are listed in Table~\ref{table}.  
\begin{table}[h]
\begin{center}
\begin{tabular}{|c|cccccccccc|}
\hline 
$f$ & $u$ & $u$ & $u$ & $u$ & $u$ & $u$ & $s$ & $u$ & $u$ & $s$ \\ 
$eB/m_\pi^2$ & 1 & 1 & 1 & 1 & 1 & 1 & 1 & 15 & 15 & 15\\
$T$, GeV & 0.2 & 0.2 & 0.2 & 0.2 & 0.2 & 0.2 & 0.2 & 0.4 & 0.4 & 0.4\\
$\phi$ & $\frac{\pi}{3}$ & $\frac{\pi}{3}$ &  $\frac{\pi}{3}$ &  $\frac{\pi}{3}$ & $\frac{\pi}{6}$ & $\frac{\pi}{12}$ &$\frac{\pi}{3}$ & $\frac{\pi}{3}$ & $\frac{\pi}{3}$ & $\frac{\pi}{3}$ \\ 
$k_\bot$,~GeV & 0.1 & 1 & 2& 3 & 1 & 1& 1& 1 & 2 & 1\\
$x$ &0.096 & 9.6 & 38& 86& 29& 35& 19& 0.64& 2.6& 1.3\\ \hline
$j_\text{max}$ & 30 & 40 & 90 & 150 & 120 & 200& 90& 8 &  12& 16 \\
\hline
\end{tabular}
\end{center}
\caption{The upper summation limit in  \eq{spec1} that yields the 5\% accuracy. $j_\text{max}$ is the highest Landau level  of the initial quark that is taken into account at this accuracy. Throughout the table $y=0$.}\label{table}
\end{table}
%

\subsection{High  photon energy}

\begin{figure}[ht]
      \includegraphics[height=5cm]{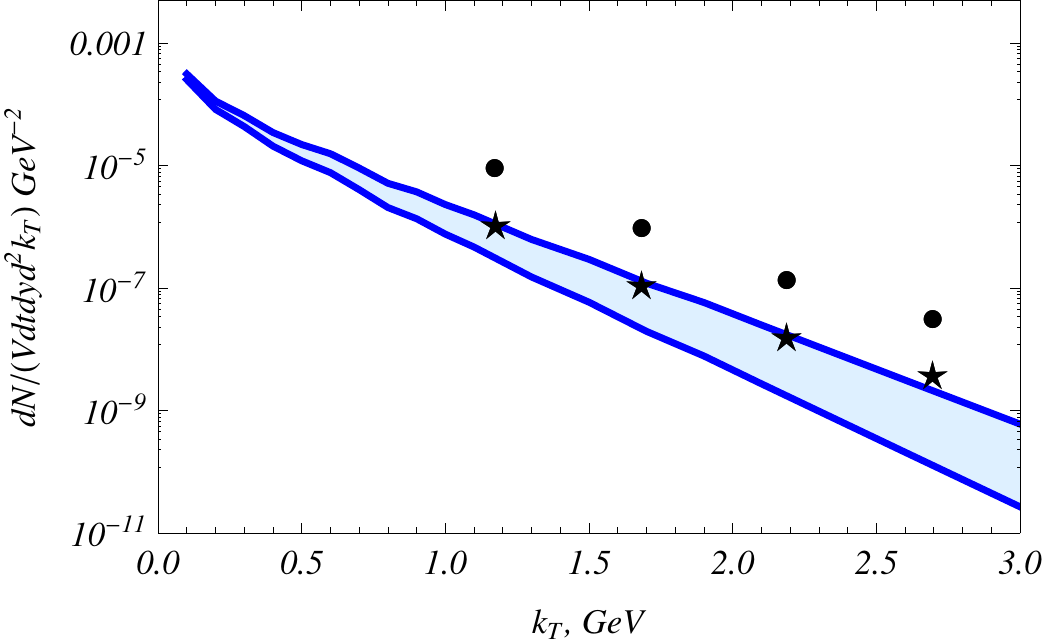} 
  \caption{Variation of the synchrotron spectrum with plasma temperature. Lower line: $T=200$~MeV, upper line: $T=250$~MeV. Other parameters are the same as in \fig{tot-synch}(a).}
\label{synch-err}
\end{figure}

The high energy tail of the photon spectrum is quasi-classical and approximately continuous. In this case the Laguerre polynomials can be approximated by the Airy functions or the corresponding  modified Bessel functions. The angular distribution of the spectrum can be found in \cite{Baring:1988a}:
\beql{hel}
\frac{dN^\text{synch}}{Vdt d\Omega d\omega }=\sum_f\frac{z_f  ^2\alpha}{\pi}\frac{n_f \omega m^2}{4T^3}\sqrt{\frac{e_fBT\sin\theta}{m^3}} e^{-\omega/T}\,,
\eeq
provided that $\omega\gg m\sqrt{mT/e_fB\sin\theta}$. Here $n_f$ is number density of flavor $f$, which is independent of $B$:
\beql{n.den}
n_f= \frac{2\cdot 2N_c\,e_fB}{4\pi^2}\sum_{j=0}^\infty\int_{-\infty}^\infty dp\, e^{-\e_j/T}\approx \frac{4N_c}{\pi^2}\,T^3\,.
\eeq
Here summation over $j$ was replaced by integration. It follows that this part of the spectrum increases with magnetic field strength  as $\sqrt{B}$ and and with temperature as $\sqrt{T}e^{-\omega/T}$. Therefore, variation of the spectrum with $T$ is much stronger than with $B$. The $T$ dependence is shown in \fig{synch-err}.

\section{Pair annihilation}\label{sec:annih}

The theory of  one-photon pair annihilation was developed in \cite{Harding:1986a,Wunner:1986a}.  It was shown in \cite{Wunner:1979a} that in the super-critical regime $eB\gg m^2$ one-photon annihilations is much larger than the two-photon annihilation.  In this section the one-photon annihilation of $q$ and $\bar q$ pairs in the QGP is calculated. 

For $q\bar q$ pair annihilation the conservation of energy and momentum is given by
\beql{conserv2}
\e_j+\e_k=\omega \,,\quad p+q= \omega\cos\theta\,.
\eeq
The spectral density of the annihilation rate per unit volume reads
\begin{align}\label{anih1}
\frac{dN^\text{annih}}{Vdtd\omega d\Omega}= \sum_f\frac{\alpha z_f^2 \omega N_c}{4\pi e_f  B}
\sum_{j=0}^\infty \sum_{k=0}^\infty \int dp\, \frac{2e_f   B}{2\pi^2}f(\e_j)
\int dq\, \frac{2e_f   B}{2\pi^2}f(\e_k)&\nonumber\\
\times \delta(p+q-\omega \cos\theta)\delta(\e_j+\e_k-\omega)\{ |\mathcal{T}_\bot|^2 + |\mathcal{T}_\parallel|^2\}\,, &
\end{align}
where the matrix elements $\mathcal{T}$ can be obtained from \eq{Mmat1},\eq{Mmat2} by making substitutions $\e_k\to -\e_k$, $q\to -q$ and  are given by
\begin{align}
4\e_j\e_k|\mathcal{T}_\bot|^2=& (\e_j\e_k-pq+m^2)[I_{j,k-1}^2+I_{j-1,k}^2]-2\sqrt{2j e_f   B}\sqrt{2k e_f   B}[I_{j,k-1}I_{j-1,k}]\label{Tmat1}\,.\\
 4\e_j\e_k|\mathcal{T}_\parallel|^2=& \cos^2\theta\big\{ ( \e_j\e_k-pq+m^2)[I_{j,k-1}^2+I_{j-1,k}^2]+2\sqrt{2je_f   B}\sqrt{2k e_f   B}[I_{j,k-1}I_{j-1,k}]\big\}\nonumber\\
 &-2\cos\theta\sin\theta\big\{ -p\sqrt{2 k e_f   B}[I_{j-1,k}I_{j-1,k-1}+I_{j,k-1}I_{j,k}]
 \nonumber\\
 &+q\sqrt{2je_f   B}[I_{j,k}I_{j-1,k}+I_{j-1,k-1}I_{j,k-1}]\big\}\nonumber\\
 &+ \sin^2\theta\big\{ (\e_j\e_k+pq+m^2)[I_{j-1,k-1}^2+I_{j,k}^2]-2\sqrt{2je_f   B}\sqrt{2ke_f   B}(I_{j-1,k-1}I_{j,k})\big\}\,, \label{Tmat2}
\end{align}
with the same functions $I_{i,j}$ as in \eq{ijk}. Integration over $q$ removes the  delta function responsible for the conservation of momentum along the field direction. The remaining delta function is responsible for  energy conservation and can be written in exactly the same form as in \eq{re1} with particle energies and momenta now obeying the conservation laws \eq{conserv2}.  It is straightforward to see that  momentum $p^*_\pm$ is still given by \eq{p*},\eq{mm}. The photon spectrum produced  by annihilation of quark in state $j$ with antiquark in state $k$ has a threshold 
$\omega_{a,ij}$ that is given by the  case (ii) in \eq{cases}:
\beql{cuta}
\omega\ge \omega_{a,ij}=\frac{m_j+m_k}{\sin\theta}= \frac{\sqrt{m^2+2j e_f   B}+\sqrt{m^2+2ke_f   B}}{\sin\theta}\,.
\eeq
Thus, the spectral density of the annihilation rate per unit volume is 
\begin{align}\label{spec2}
\frac{dN^\text{annih}}{Vdt d\omega d\Omega}= \sum_f\frac{\alpha z_f^2 \omega N_c}{4\pi^5}e_f  B 
\sum_{j=0}^\infty \sum_{k=0}^\infty 
\vartheta(\omega-\omega_{a,ij})
\int dp \sum_\pm\frac{\delta(p-p^*_\pm)}
{\big| \frac{p}{\e_j}-\frac{q}{\e_k}\big|}
&\nonumber\\
\times\left\{
|\mathcal{T}_\bot|^2+|\mathcal{T}_\parallel|^2\right\}f(\e_j)f(\e_k)\,. &
\end{align}
Passing to $y$ and $p_\bot$ variables in place of $\omega$ and $\theta$ is similar to  \eq{mult2}.

\begin{figure}[ht]
\begin{tabular}{cc}
      \includegraphics[height=5cm]{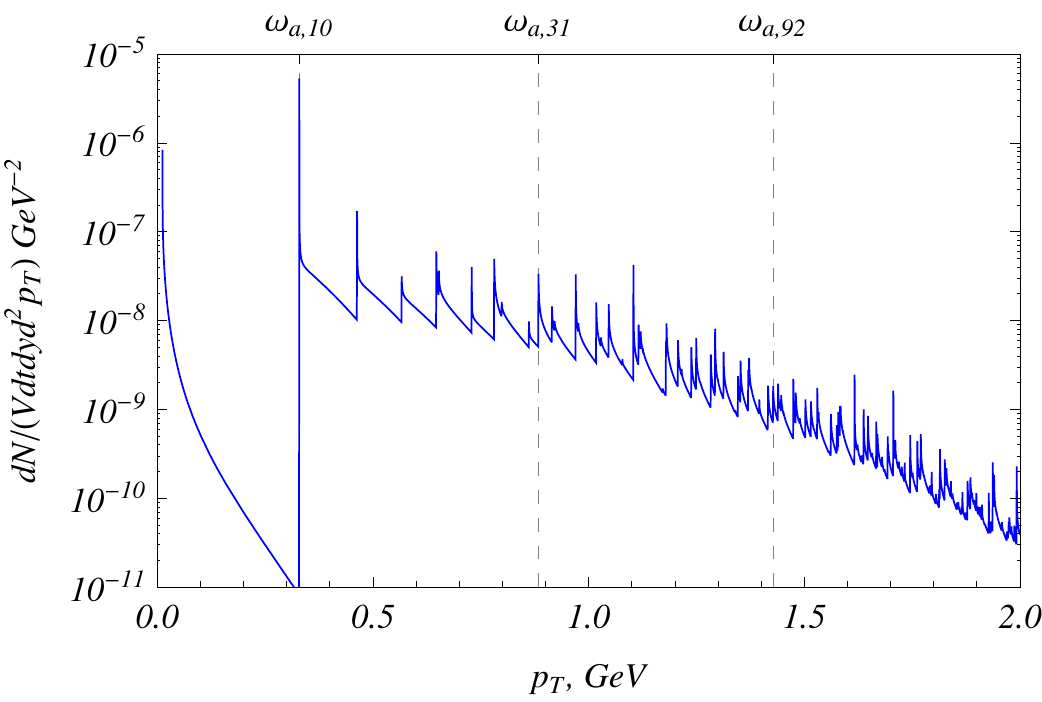} &
      \includegraphics[height=5.cm]{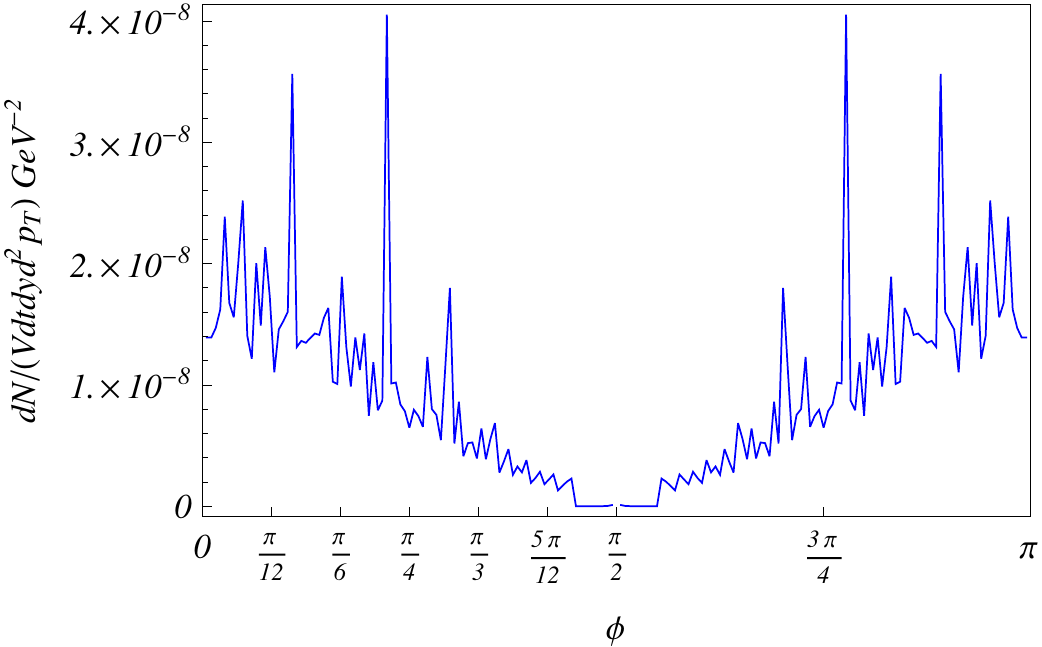}\\
      $(a)$ & $(b)$ 
      \end{tabular}
  \caption{ Photon spectrum in one-photon annihilation of $u$ and $\bar u$ quarks. $eB= m_\pi^2$, $y=0$. (a) $k_\bot$-spectrum  at  $\phi=\pi/3$, (b) azimuthal angule distribution at $k_\bot=1$~GeV.  }
\label{annih}
\end{figure}

The results of the numerical calculations are represented in \fig{annih}. Panel (a) shows the spectrum of photons radiated in annihilation of $u$ and $\bar u$. We conclude that contribution of the annihilation channel is negligible as compared to the synchrotron radiation. 

\section{Conclusions}\label{sec:concl}

Results of the calculations performed in this article indicate that photon production by QGP due to its interaction with external magnetic field give a considerable contribution to the total photon multiplicity in heavy-ion collisions. This is seen in \fig{tot-synch} were the model calculation is compared with the experimental data \cite{Adare:2008ab}.   The two processes were considered: synchrotron radiation and pair annihilation. In the kinematic region relevant for the current high energy heavy-ion experiments,  contribution of the synchrotron radiation is about two orders of magnitude larger than that of pair annihilation. The largest contribution to the photon multiplicity arises from photon momenta of the order of $\sqrt{eB}$. This may provide an explanation of the photon excess  observed by the PHENIX experiment \cite{Adare:2008ab}. Similar mechanism is also responsible for enhancement 
of low mass di-lepton production that proceeds via emission of virtual photon which subsequently decays into di-lepton pair.

One possible way to ascertain the contribution of electromagnetic radiation in external magnetic field is to isolate the azimuthally symmetric component with respect to the direction of the magnetic field. It seems that synchrotron radiation dominates the photon spectrum at low $k_\bot$. Thus, azimuthal symmetry can be easily checked by simply plotting the multiplicity vs $\omega$, $\theta$ and $\varphi$, where $\omega$ is photon energy, $\theta$ is emission angle with respect to the magnetic field and $\varphi$ is azimuthal angle around the magnetic field direction (which is perpendicular both to the collision axis and to the impact parameter). In \fig{synch}(a) it is also seen that in these variables it may be possible to discern the cutoff frequencies $\omega_{s,jk}$ that appear as resonances (in  \fig{synch} $y=0$ so $k_\bot = \omega$). Note that averaging over the azimuthal angle $\alpha$ around the collision axis direction distroys these features, see \fig{tot-synch}. 

The greatest source of uncertainty in the present calculation stems from treating the magnetic field as constant. It is inevitable that it has spatial \cite{Bzdak:2011yy} and temporal variations, which will modify the photon spectrum. Analytical calculations of these effects present a serious challenge, but may be tackled in the quasi-classical approximation. Novel computational techniques, such as discussed in \cite{Zhao:2011ct} is another promising avenue for investigating the particle production in external fields.

\acknowledgments
I thank  Yoshimasa Hidaka and Kazunori Itakura for useful  correspondence and James Vary for discussions of related topics. 
This work  was supported in part by the U.S. Department of Energy under Grant No.\ DE-FG02-87ER40371.

\appendix
\section{Magnetic component of the QGP energy density }\label{appA}

Energy density associated with magnetic field is (in Gauss units)
\beql{mag.f}
\epsilon_M=(eB)^2/(8\pi \alpha)\,.
\eeq
Assume that the effect of magnetic field on the QGP is weak and its energy density is much smaller than the total QGP energy density $\epsilon_M\ll \epsilon_\text{QGP}$. Then energy density of ideal QGP  containing $N_f$ quark flavors at temperature $T$ is given by 
\beql{en.id}
\epsilon_\text{QGP}=\frac{\pi^2}{30}T^4\left(2(N_c^2-1)+\frac{7}{8}4N_cN_f\right).
\eeq
With $N_c=3$, $N_f=2$ we arrive at the following ratio
\beql{rat.mag}
\frac{\epsilon_M}{\epsilon_\text{QGP}}= 0.45 \frac{(eB)^2}{T^4}\,.
\eeq
At RHIC $eB\sim m_\pi^2$ and $T\sim 2 m_\pi$, so that at early times about 3\% of energy density of plasma resides in the magnetic field. At LHC, $eB\sim 15m_\pi^2$ and $T\sim 4m_\pi$ so that as much as 40\% is stored in magnetic field! 
This signals that magnetic field plays a crucial role in QGP dynamics at LHC energies. 



\end{document}